\documentclass[letterpaper, 10 pt, conference]{ieeeconf}  

\IEEEoverridecommandlockouts                              

\IEEEoverridecommandlockouts                              
\usepackage[none]{hyphenat}
\usepackage{amsmath}
\usepackage{amssymb}
\usepackage{amsfonts}
\usepackage{amstext}
\usepackage{mathrsfs}
\usepackage{amsthm}
\usepackage{cite}
\usepackage{graphicx}
\usepackage{float}
\usepackage{comment}
\usepackage{fullpage}
\usepackage{nomencl}
\usepackage[capitalise]{cleveref}
\usepackage{mathtools}

\usepackage{color}

\usepackage{tikz}
\usetikzlibrary{positioning}
\tikzset{block/.style = {draw, rectangle},
Connect/.style={-latex,thick}}
\usetikzlibrary{shapes,snakes}
\usetikzlibrary{calc}

\newtheorem{theorem}{Theorem}[section]
\theoremstyle{remark}

\theoremstyle{definition}

\newtheorem{prop}{Proposition}[section]

\newtheorem{assumption}{Assumption}[section]

\title{\LARGE \bf Delay-Induced Watermarking for\\  Detection of Replay Attacks in Linear Systems}

\author{Christoforos Somarakis$^{1}$, Raman Goyal$^{2}$, Erfaun Noorani$^{3}$, Shantanu Rane$^{2}$
\thanks{$^{1}$ Christoforos Somarakis is with the Data Science and Applied Mathematics Group, Merck \& Co, West Point, PA, USA
        {\tt\small christoforos.somarakis@merck.com}}%
\thanks{$^{2}$ Raman Goyal, Shantanu Rane are with Palo Alto Research Center - An SRI Company, Palo Alto, CA, USA  {\tt\small \{rgoyal,srane\}@parc.com}}
\thanks{$^{3}$ Erfaun Noorani is with the Department of Electrical and Computer Engineering, the Institute for System Research (ISR) at the University of Maryland College Park, College Park, MD, USA. E. Noorani is a Clark doctoral Fellow at A. James Clark School of Engineering.
         {\tt\small enoorani@umd.edu}}%
}

\begin{document}

\maketitle
\thispagestyle{empty}
\pagestyle{empty}

\begin{abstract}
A state-feedback watermarking signal design for the detection of replay attacks in linear systems is proposed. The control input is augmented with a random time-delayed term of the system state estimate, in order to secure the system against attacks of replay type. We outline the basic analysis of the closed-loop response of the state-feedback watermarking in a LQG controlled system. Our theoretical results are applied on a temperature process control example. While the proposed secure control scheme requires very involved analysis, it, nevertheless, holds promise of being superior to conventional, feed-forward, watermarking schemes, in both its ability to detect attacks as well as the secured system performance. 
 \end{abstract}

\section{INTRODUCTION}

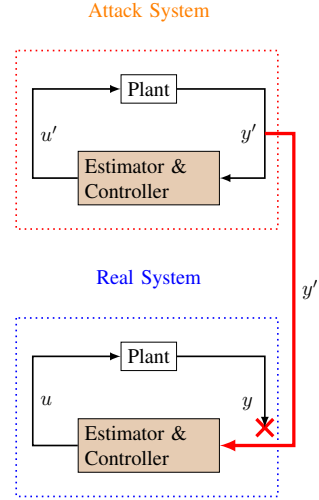
\begin{figure}
\centering
\resizebox{120 pt}{190 pt}{%
\begin{tikzpicture}
    \tikzstyle{ann} = [draw=none,fill=none]
    \node [block] (virtualplant) {Plant};
     \node [ann, right of=virtualplant, node distance = 2cm](nodeaux1) {};
      \node [ann, left of=virtualplant, node distance = 2cm](nodeaux2) {};
      \node [ann, below of=nodeaux1, node distance = 0.75cm](nodeaux5) {};
          \node [ann, right of=nodeaux5, node distance = 0.5cm](nodeaux6) {};
    \node [block, fill=brown!40, below of=virtualplant, node distance=1.5cm, text width=2.2cm ] (virtualcontrol) {Estimator \& Controller};
    \node [ann, right of=virtualcontrol, node distance = 2cm](nodeaux3) {};
     \node [ann, left of=virtualcontrol, node distance = 2cm](nodeaux4) {};
    \draw [Connect] (virtualplant) -- (virtualplant-|nodeaux1) -- node[left]{$y'$} (nodeaux1|-nodeaux3) -- (virtualcontrol);
    \draw [Connect] (virtualcontrol) -- (virtualcontrol-|nodeaux4) -- node[right]{$u'$} (nodeaux4|-nodeaux2) -- (virtualplant);
    \draw[red,thick,dotted]($(virtualplant.north west)+(-1.8,0.4)$)  rectangle  ($(virtualcontrol.south east)+(1,-0.4)$) ;
    \node [ann, above of = virtualplant, node distance=1.3cm](label) {\textcolor{orange}{ Attack System}};
    \node[block] at (0, -4.5)  (realplant)    {Plant};
    \node[block, fill=brown!40, below of=realplant, node distance=1.5cm, text width=2.2cm ]  (realcontrol)    {Estimator \& Controller};
      \draw[blue,thick,dotted]($(realplant.north west)+(-1.8,0.4)$)  rectangle  ($(realcontrol.south east)+(1,-0.4)$) ;
    \node [ann, right of=realcontrol, node distance = 2.5cm](nodeaux7) {};
   \draw [Connect,red, ultra thick] (nodeaux5|-nodeaux6) -- (nodeaux5-|nodeaux6) -- node[right]{\textcolor{black}{$y'$}}  (nodeaux6|-nodeaux7) -- (realcontrol);
    \node [ann, left of=realplant, node distance = 2cm](nodeaux8) {};
     \node [ann, left of=realcontrol, node distance = 2cm](nodeaux9) {};
     \draw [Connect] (realcontrol) -- (realcontrol-|nodeaux9) -- node[right]{$u$} (nodeaux9|-nodeaux8) -- (realplant);
     \node[ann,right of = realplant, node distance=2cm](coaux){};
     
     \matrix[nodes={draw, ultra thick, red, fill=blue!20}, below of = coaux, node distance=1.2cm] {
    \node[cross out](co) {};\\
    };
    \node[ann,above of = co, node distance=0.4cm](coaux2){};
    \node[ann,left of = coaux2, node distance=0.3cm](coaux3){$y$};
    \draw[Connect](realplant) --(realplant-|coaux) --  (coaux|-co); 
     \node [ann, above of = realplant, node distance=1.3cm](label) {\textcolor{blue}{Real System}};
\end{tikzpicture}
}
\caption{The replay attack cast as a drive-response system. The attack signal, $y'$, intervenes and compromises the real-time measurements of the real system. Signal attack, $y'$, a recorded part of $y$, is considered as an output of a copy system running in parallel with the real system.}
\label{fig:masterslave} \vspace{-0.2in}
 \end{figure}

The widespread integration of the physical and the cyber layers within the Industry 4.0 transformation has fueled research efforts in the field of Cyber-Physical Systems security (CPSs) \cite{VENKATASUBRAMANIAN2003293}. The complex nature of CPSs makes such systems vulnerable to malicious attacks. 
A popular example of system vulnerability in a CPS environment, is this of \textit{replay attack}. In a replay attack, the attacker records the output signal of the real system and then replays it back repeatedly for the controller \cite{SurveyReplay}.  In the absence of secure control, the victim stays oblivious to the attack. Figure \ref{fig:masterslave} illustrates a system theory point of view of replay attacks: A master-slave hierarchy of identical systems with the output of the attack system to compromise the real system.  

Fast and reliable methods of detecting attacks are, therefore, of deep interest in the field of estimation and control \cite{sharma2010sensor,cardenas2008secure}. Our work focuses on a secure method based on \textit{input watermarking}. These are untraceable and unrepeatable signals, e.g. Gaussian noise, that typically get combined with the nominal input signal \cite{rubio2017use}. Their unique characteristic is that they enable detection of replay attacks whenever certain statistical analysis tools are utilized, such as the $\chi^2$-based ones \cite{mehra1971innovations}. The rate at which replay attacks are detected, as well as, the level of performance degradation of the controlled system under watermarking signal inputs, are two metrics that characterize the overall efficiency of watermarking as a secure control method \cite{mo2009secure}.
We present a novel type of watermarking, based on random time-delayed state-feedback. We analyze the resulting dynamics and discuss an example where this design can offer security with comparable, if not improved, detection rate and performance degradation as compared to mainstream watermarking that assume an additive Gaussian signal. The remainder of the paper is organized as follows: Notations are presented in \S\ref{section: notation}. The problem formulation is put in \S\ref{section: formulation}. In \S\ref{sec:preliminary}, we discuss preliminary results and in \S\ref{sec:attack} we detail system response and key statistics in presence of a replay attack. In \S\ref{sec: time-delayperformance}, we discuss the effect of feedback time-delay control in the system performance. In \S\ref{sec:sim}, we conduct an experimental study of the effectiveness of our watermarking design and benchmark our design against widely known Gaussian watermarking. The numerical example is an Industrial Engineering benchmark of temperature control. Concluding remarks are placed \S~\ref{sec: discussion}. Proofs of technical results are omitted due to space limitations.

\section{Nomenclature}\label{section: notation} 
The $n$-dimensional vector space is denoted by $\mathbb R^{n}$. It consists of vector elements $x\in \mathbb R^{n}$. The $n\times m$ zero matrix is $O_{n\times m}$, while by $O_n$ is the $n\times n$ square zero matrix. The square identity matrix is denoted by $I_n$. Analysis will leverage four state-space representations: The system state-space embedded in $\mathbb R^{n}$, the system-controller state-space embedded in $\mathbb R^{2n}$, the drive-response space embedded in $\mathbb R^{4n}$, and the uplifted drive-response space embedded in $\mathbb R^{4n(\overline{\tau}+1)}$. Projection operators are denoted by $\mathfrak{P}\big[\cdot\big]\in \mathbb R^{n}\rightarrow \mathbb R^{m}$, for $m<n$.  Notation $\mathcal N\left(\mu,\Sigma\right)$ denotes the $n$-dimensional normal distribution with mean $\mu$ and covariance matrix $\Sigma$. Symbol $\mathbb E\left[ \cdot \right]$ is reserved for the expectation operator. $\mathbb E_{\sim p}$ denotes expectation with respect to the $p$ random variable. By $\rho(A)$ we denote the spectral radius of matrix $A$. The Variance $\operatorname{Var}(\cdot)$, Covariance $\operatorname{Cov}(\cdot)$ vectorization, $\operatorname{vec}(\cdot)$ and trace $\operatorname{trace}(\cdot)$ operators are used in the standard manner.

\section{Problem Formulation}\label{section: formulation}
The plant is a discrete LTI stochastic system. The state estimator is a Kalman filter providing the plant state estimate. The state estimator feedback optimizes a linear quadratic cost by regulating output towards zero. Lastly, another feedback control comprising the watermarking signal is added together with the optimal LQG feedback.  
\subsection{Real System Dynamics}
\subsubsection{Plant Equations} The state of plant $x\in \mathbb R^{n_x}$ evolves in time $t \in \mathbb Z_+$ according to:  
\begin{align}
x_{t+1} &= A\, x_t +B\, u_t +  w_t,\label{state_eqn2} \\
y_t &= C \, x_t + v_t,  \label{output_eqn2}
\end{align}
where $x_t \in \mathbb{R}^{n_x}$ is the state of the system and $u_t \in \mathbb{R}^{n_u}$ is the control input, at time $t$. Process noise $w_t \sim \mathcal{N}(\mathbf {0},\Sigma_{W})$, $\forall\,t$ and measurement noises $v_t \sim \mathcal{N}(\mathbf {0},\Sigma_{V})$, $\forall\, t$ follow zero mean Gaussian distribution with respective covariances. 
\subsubsection{Controller Equations}
State vector is regulated around a reference value in the mean sense, with a controller that optimizes some cost function.
\paragraph{Kalman Filter} For the predicted state estimate we use the notation $\hat{x}_t=x_{t|t-1}$ and for the updated state the notation $\tilde{x}_t=x_{t|t}$. The state dynamics are:
\begin{equation}\label{eq: kalman}
    x_{t+1|t}=A\, x_{t|t-1} + B\, u_t + L\,(y_t - C\, x_{t|t-1}),
\end{equation} where $L:=APC^T (CPC^T+\Sigma_V)^{-1}$ is the $n_x\times n_y$ Kalman gain and $P$ is the $n_x\times n_x$ matrix solution of the algebraic Ricatti equation $P=A\big[ P - P C^T (C P C^T + \Sigma_V)^{-1}CP\big] A^T + D \Sigma_{W} D^T.$  The Kalman output is 
\begin{equation}\label{eq: kalmanoutput}
x_{t|t}  = (I_{n_x}-MC)\,x_{t|t-1} + M y_t,
\end{equation} with $M=PC^T(CPC^T+\Sigma_{V})^{-1}$ the  innovation gain. 
\paragraph{LQG Controller} We implement the feedback
\begin{equation}\label{eq: controlformlqg}
    u_{t}^{\text{LQG}} = - K_{\tilde{x}}\, \tilde{x}_{t}, 
\end{equation}
where gain $K_{x}$ is designed to minimize the cost:
\begin{equation}\label{eq: lqg}
    J = \lim_{t\rightarrow \infty} \frac{1}{t} \mathbb E \bigg[\sum_{k=0}^{t-1} x_k^T\,Q\, x_k + u_k^T\, R\, u_k \bigg].
\end{equation}
It is well-known,\cite{astrom}, that $K_{\tilde{x}} =  (R+B^T \,\Delta\, B)^{-1}  B^T  \Delta A$ with $\Delta$ solves $ \Delta = Q + A^T \Delta\, A - A^T \Delta \,B (R+B^T\Delta\,B)^{-1}B^T\Delta\, A $. 
\paragraph{Watermarking} We utilize a delayed version of state estimate with a time-varying delay with the form
\begin{equation}\label{eq: controlformdelay}
    u_t^{\text{WM}} = - K_{\tau}\, \tilde{x}_{t-\tau_t},
\end{equation} where the gain matrix $K_{\tau}$, and the time-delay $\tau_t$, are design parameters. We consider time-delays as random variables such that for every $t$, $\tau_t\in \{1,\dots,\overline{\tau}\}$ with mass probability function $p_{\tau}$. Time series $\{\tau_t\}_{t \geq 0}$ consists of IID random variables that are also independent of plant process and sensor noises. In conclusion, the control in our problem has the form 
\begin{equation}\label{eq: controlform}
    u_{t} = u_{t}^{\text{LQG}} +   u_t^{\text{WM}}
\end{equation}
with $u_{t}^{\text{LQG}}$ as in \cref{eq: controlformlqg} and $ u_t^{\text{WM}}$ as in \cref{eq: controlformdelay}.
 
\subsection{System-Controller Dynamics} The coupled dynamics enable the augmented state $\boldsymbol{x}_t := \big( x_t^T , \hat{x}_t^T \big)^T \in\mathbb R^{2n_x}$, $\boldsymbol{n}_t : = \big( w_t^T , v_t^T , v_{t-\tau_t}^T \big)^T \in \mathbb R^{n_x+2n_y},$ where $\hat{x}_t : = x_{t|t-1}$ from \cref{eq: kalman}, evolving as
\begin{equation}\label{eq: clsystem}
\boldsymbol{x}_{t+1}= \mathbf A \, \boldsymbol{x}_{t} + \mathbf B \, \boldsymbol{x}_{t-\tau_t} + \mathbf{\Gamma}\, \boldsymbol{n}_t, ~~~t>\overline{\tau},
\end{equation}
for appropriate $\mathbf A \in \mathbb R^{2n_x\times 2n_x}$, $\mathbf B\in \mathbb R^{2n_x\times 2n_x}$, $\mathbf{\Gamma} \in \mathbb R^{2n_x\times n_x+2n_y}$.
\subsection{Attack System Dynamics}
Replay attacks take place with an attacker compromising the real system sensing. The attack time $t'$ is defined as the first time instance when sensor component, instead of $y_t$, transmits measurement signal $y_{t}'$, for $t\geq t'$. Compromised signal $y'=\{y'_t\}_{t=t_{\text{start}}}^{t_{\text{end}}}$ is assumed to be a signal of the true output \cref{output_eqn2} recorded by the attacker at past time, and being replayed during the attack in a loop. Following the line of arguments as in \cite{mo2009secure}, we assume that $y'_t$ is the result of a \textit{virtual} system, i.e. a copy of \cref{state_eqn2,output_eqn2,eq: kalman,eq: kalmanoutput} with control \cref{eq: controlform} that is operating in parallel with the real system. The attack system dynamics are
\begin{equation}\label{eq: mastersystem}
    \begin{split}
         x'_{t+1} & = A\, x'_{t} + B\, u'_{t} + w'_t \\
          y'_t & = C\, x'_t + v'_t\\
          \hat{x}_{t+1}' & =  A\,\hat{x}_t' + B\,u_t' + L\,(y_t' - C\,\hat{x}_{t}')
    \end{split}
\end{equation}
The control law is $
   u_t'=-K_{\tilde{x}}(I-MC)\,\hat{x}_{t}'-K_{\tilde{x}}MC\,x_{t}'-K_{\tilde{x}}M\,v_t' - K_{\tau}(I-MC)\,\hat{x}_{t-\tau_t'}'-K_{\tau}MC\,x_{t-\tau_t'}'-K_{\tau}M\,v_{t-\tau_t'}'.$ 
When the attack moment, $t'$, is large enough, the sources of randomness in the attack system can be assumed independent from the real system's ones.
\subsection{Drive-Response System Under Replay Attack}
During a replay attack the real system measures a compromised signal, e.g. a past copy of the real system. This signal could be provided by \eqref{eq: mastersystem}. For implementation purposes, the attack signal occurs as an information leak of measurement for a considerable amount of time, say leaked data set $\{y_{t}\}_{t=t_1}^{t_2}$, so that replay signal $y'$ is a repeat in a loop. The real system (plant and estimator) operates  then according to 
\begin{equation}\label{eq: slavesystem}
\begin{split}
    x_{t+1} & = A\,x_t + B\,u_t + w_t  \\
    y_t &  = C\,x_t+v_t \\
    \hat{x}_{t+1} & =  A\,\hat{x}_t + B\,u_t + L\,(y_t' - C\,\hat{x}_{t})
\end{split}
\end{equation} and now
the control law \eqref{eq: controlform} takes the form $   u_t=-K_{\tilde{x}}(I-MC)\,\hat{x}_{t}-K_{\tilde{x}}MC\,x_{t}'-K_{\tilde{x}}M\,v_t' - K_{\tau}(I-MC)\,\hat{x}_{t-\tau_t}-K_{\tau}MC\,x_{t-\tau_t}'-K_{\tau}M\,v_{t-\tau_t}'.$ One can cast \eqref{eq: mastersystem} and \eqref{eq: slavesystem} as a drive-response system architecture where the attack system \eqref{eq: mastersystem} drives the dynamics of \eqref{eq: slavesystem}. We construct the drive-response state-space vectors:  
$\boldsymbol x_t^{\dagger} := \big( x_t^T, \hat{x}_t^T, x_t'^T, \hat{x}_t'^T\big)^T \in \mathbb R^{4n_{x}}$ $\boldsymbol{n}_{t}^{\dagger}:=\big(
 w_t^T, w_t'^T, v_t'^T, v_{t-\tau_t}'^T, v_{t-\tau_t'}'^T \big)^T   \in \mathbb R^{2n_{x}+3n_{y}},$
with associated dynamics to be 
\begin{equation}\label{eq: drive-response}
 \boldsymbol x_{t+1}^{\dagger} = \mathbf{A}^\ddagger\,\boldsymbol x_t^{\dagger} + \mathbf B^\ddagger\,\boldsymbol x_{t-\tau_t}^{\dagger} +   \mathbf C^\ddagger\,\boldsymbol x_{t-\tau'_t}^{\dagger} + \mathbf G^\ddagger\,\boldsymbol{n}_t^{\dagger} 
\end{equation}
with matrices $ \mathbf A^\ddagger,  \mathbf B^\ddagger,  \mathbf C^\ddagger,  \mathbf G^\ddagger$ of appropriate form. Note that unless a replay-attack takes place, $\boldsymbol{x}_{t}^\dagger$ represents the state of two decoupled identical systems. 
\subsection{Uplifted Drive-Response Level Dynamics} The core difficulty in dynamics \eqref{eq: drive-response} is the presence of time-delays that also vary with time hindering an explicit expression of solution to be used in the attack detection. One way to deal with this challenge is by grouping together $\boldsymbol{x}_{t}^{t-\overline{\tau}}$. The uplifted vectors are $\mathbb X_{t}= \big[ \boldsymbol{x}_{t}^\dagger , \boldsymbol{x}_{t-1}^\dagger, \dots, \boldsymbol{x}_{t-\overline{\tau}}^\dagger \big] \, \in \mathbb R^{4n_{x}(\overline{\tau}+1)}$ and $\mathbb N_{t} \leftrightarrow \boldsymbol{n}_{t}^{\dagger}.$ The dynamics in the uplifted space read as:
\begin{equation}\label{eq: uplifteddriveresponsedynamics}
    \mathbb X_{t+1} = \mathcal A_t\, \mathbb X_{t} + \mathcal G\,\mathbb N_{t}
\end{equation}
with $\mathcal A_{t}=\mathcal A_{\tau_t,\tau^{'}_t}$ a time-dependent (random) matrix, and $\mathcal G$ a constant matrix.
For every $t>0$, $\mathcal A_t=\mathcal A_{\tau_t,\tau^{'}_t}$ is sampled according to probability mass function $\{p_{\tau}p_{\tau'}\}_{1}^{\overline{\tau}}$ with 
$\overline{\mathcal A}:=\mathbb E\left[ \mathcal A_t\right] =\mathbb E\left[ \mathcal A_{\tau_t,\tau^{'}_{t}} \right] = \sum_{\tau, \tau'=1}^{\overline{\tau}} \mathcal A_{\tau,\tau'}\,p_{\tau}\,p_{\tau'}$
to be the expected value of $\mathcal A$. Turning system \cref{eq: clsystem} into \cref{eq: uplifteddriveresponsedynamics} suppresses time-delays at the expense of embedding the solution in a high-dimensional space evolving with time-dependent (random) dynamics. The vector $\mathcal G\, \mathbb N_t$ involves random variables distributed according to $\{p_\tau\}$, $\{p_\tau'\}$ as well as normal distribution. Solution of \cref{eq: uplifteddriveresponsedynamics} can be represented for $t>\overline{\tau}$ as
\begin{equation}\label{eq: randommatricessolution}
    {\mathbb X}_{t} = {\mathcal A}_{t:0}\,{\mathbb X}_0 + \sum_{k=0}^{t-1} {\mathcal A}_{t: k+1}\, {\mathcal G}\, {\mathbb N}_{k}
\end{equation} where $\mathcal A_{t_2:t_1} := \mathcal{A}_{t_2-1}\cdots \mathcal{A}_{t_1}$ for $t_2>t_1$ is the transition matrix, with the standard properties: $ \mathcal A_{t_2,t_1} = \mathcal{A}_{t_2,t'} \mathcal A_{t',t_1},\, \forall\, t_2\geq t' \geq t_1 $,
 and $\mathcal A_{t:t}=I_{4n_x(\overline{\tau}+1)}$. Evidently, $\mathcal A_{t_2:t_1}$ is a product of $t_2-t_1$ independent and identically distributed matrix-valued random variables each of which encapsulates the (mutually independent) random variables $\tau_t$ and $\tau'_t$ from the real and the attack system accordingly. 
\section{Preliminaries} \label{sec:preliminary}  The stability of the free dynamics of \cref{eq: clsystem} characterize the long-term properties of the corresponding state matrices in \eqref{eq: drive-response} and \eqref{eq: uplifteddriveresponsedynamics}.
\subsection{Stability on the System-Controller Level Dynamics} Watermarking signals are expected to impact negatively the overall system performance. The proposed watermarking as a feedback loop with time-delays may also destabilize the real system. It is therefore critical to design feedback gain matrices $K_{\tau}$ for $u_{t}^{WM}$ in \eqref{eq: controlformdelay} and distributions $\{1,\dots,\overline{\tau}\}$ such that $\mathbf{x}_t:= \mathbb E_{\sim w,v}[\boldsymbol{x}_t]$ satisfies
$\lim_{t\rightarrow +\infty}  \mathbf{x}_t:=\lim_{t\rightarrow +\infty}   \mathbb E_{\sim w,v}[\boldsymbol{x}_t] = 0.$  The result of this section states sufficient conditions for existence of such $K_{\tau}$ and $\overline{\tau}$. There are a few approaches we can adopt towards this end. Without loss of generality, we assume $\tau_t\in \{1,\dots,\overline{\tau}\}$ to be some deterministic function of time taking values in $\{1,\dots,\overline{\tau}\}$. These dynamics  satisfy:
\begin{equation}\label{eq: clsystem_mean}
    \mathbf{x}_{t+1}=\mathbf A \,  \mathbf{x}_t + \mathbf B \,  \mathbf{x}_{t-\tau_t}.
\end{equation}
%
%
The LQG control makes $\mathbf A$ Hurwitz, so that the Lyapunov equation $ \mathbf A^T \, \mathbf H \, \mathbf A - \mathbf H = - \mathbf C$ has a unique positive definite solution $\mathbf H=\mathbf H^T$, for any positive definite $\mathbf C =\mathbf C^T$. 
%
%
Let $c>0$ be the smallest eigenvalue of $\mathbf C$ and $0<\underline{\eta}\leq \overline{\eta}$ the largest and the smallest eigenvalues of $\mathbf H$. 
Define 
\begin{align}
        \alpha &: = \begin{cases} \frac{\big|\overline{\eta}-c + |\mathbf A^T \mathbf H \mathbf B | \big| }{\overline{\eta}},  & c>|\mathbf A^T \mathbf H \mathbf B| \\
      \frac{\big| \underline{\eta}-c + |\mathbf A^T \mathbf H \mathbf B |\big|}{\underline{\eta}},   &c\leq |\mathbf A^T \mathbf H \mathbf B|
    \end{cases}\label{eq: alpha}, \\
\beta&:= \underline{\eta}^{-1} \big(|\mathbf A^T \mathbf H \mathbf B|+ |\mathbf B^T \mathbf H \mathbf B|\label{eq: beta}\big).
 \end{align} 
 
\begin{theorem}\label{thm: stability} Let \cref{eq: clsystem_mean} with $\tau_t:[0,\infty)\rightarrow \mathcal T$ for $\mathcal T\subset \mathbb N$, and $\overline{\tau}=\max_{\tau \in \mathcal T} \{ \tau \} < \infty$, and its solution $\mathbf{x}$. Assume that $\alpha+\beta < 1 $ for $\alpha,\beta$ as in \cref{eq: alpha} and \cref{eq: beta}. Then $\mathbf x$ converges to zero exponentially fast.
\end{theorem}
 The probabilistic counterpart of the long term behavior of $\mathbf x_t$ follows naturally. Under conditions of Theorem \ref{thm: stability}, $\{\mathbf{x}_t\}_t$ converges to 0, almost surely and in probability. Furthermore, $\|\mathbf{x}_t\|$ is bounded hence $\{\mathbf x_t\}_t$ converges with respect to the $L^r$ norm for every $r>0$. 

\subsection{Stability on the Drive-Response Level Dynamics} In the event of an attack, the virtual system remains unaffected as it drives the real system. The latter's internal stability gets compromised, leading to performance degradation or instability. The case of interest is the former one.
\begin{assumption}\label{assum: driverresponsestability}
The drive-response system under attack, is internally asymptotically stable.
\end{assumption}
\subsection{Stability on the Uplifted Level Dynamics}
Theorem \ref{thm: stability} together with Assumption \ref{assum: driverresponsestability} leads on conclusions about the long term behavior of $\mathcal A_t$ and the quantity $\mathbb A:= \mathbb E_{\sim \tau, \tau'}\big[ \mathcal A_{\tau,\tau'} \otimes \mathcal A_{\tau,\tau'} \big]$ that will come of use in the next section.
\begin{prop}\label{prop: auxstability} The following conditions hold true:
\begin{enumerate}
    \item $ \lim_{t\rightarrow +\infty }\mathcal{A}_{t:t_0} = O_{4n_x(\overline{\tau}+1)}$ almost surely. 
    \item  $\rho\big(\mathbb A \big)<1$.
\end{enumerate}
\end{prop}
\subsection{The Asymptotic Auto-Covariance} A central quantity to this work is this of the asymptotic covariance matrix of $\boldsymbol{x}_t$ which we can express via $\mathbb X_t$ and appropriate projection. Let $\mathcal{C}=\lim_{t\rightarrow +\infty}\operatorname{Cov}(\mathbb X_t, \mathbb X_t)$ which is equal to \begin{equation}
    \mathcal{C}=\lim_{t\rightarrow +\infty} \mathbb E \big[\mathbb X_{t} \mathbb X_{t}^T \big]
\end{equation} The derivation of a formula of $\mathcal C$ involves manipulation of \eqref{eq: randommatricessolution}, i.e. calculating expectations of random products with statistically dependent quantities. The technical details are overwhelmingly complicated, though calculation remains tractable. For our purposes we only state the formula of $\mathcal C$ which is:
\begin{equation}\label{eq: ssautocovariance}
    \mathcal C = \operatorname{vec}^{-1}\bigg(  \big(I_{16n^2_x(\overline{\tau} +1)^2}-\mathbb A\big)^{-1} \boldsymbol{\omega} \bigg) 
\end{equation}
 where
 \begin{equation*}
     \begin{split}
     \boldsymbol{\omega} = & \operatorname{vec}\big(\mathcal{G} \mathcal{Q} \mathcal{G}^T \big) + \mathbb E_{\sim \tau,\tau'}\bigg[\mathcal{A}_{\tau,\tau'}  \otimes \mathcal{G} \mathcal S \mathcal{G}^T \operatorname{vec}(\overline{\mathcal{A}}^T)^{\tau-1}\bigg ] \\
     & + \mathbb E_{\sim \tau,\tau'}\bigg[ \mathcal{G} \mathcal{S} \mathcal{G}^T \otimes\mathcal{A}_{\tau,\tau'} \operatorname{vec}\big(\overline{\mathcal{A}}^{\tau-1} \big)\bigg] + \\
     & \sum_{l=1}^{\overline{\tau}} \mathbb E_{\sim \tau,\tau'}\bigg[ \big(\mathcal{A}_{\tau,\tau'} \otimes \mathcal{G}\, \mathcal{R}_{l,\tau,\tau'}\mathcal{G}^T\big) \operatorname{vec}\big( (\overline{\mathcal{A}}^{T})^{l-1} \big) \\
      & ~~~~~~~~~~ +  \big(\mathcal{G}\, (\mathcal{R}_{l,\tau,\tau'}\big)^T\mathcal{G}^T\otimes \mathcal{A}_{\tau,\tau'}\big) \operatorname{vec}\big(\overline{\mathcal{A}}^{l-1} \big) \bigg].
     \end{split}
 \end{equation*}
and matrices $\mathcal R_{(\cdot)}$, $\mathcal Q$ and $\mathcal S$ stated in Appendix \S \ref{append: matrices}.

 \section{Replay Attacks \& Detectability Analysis} \label{sec:attack}  
Our hypothesis is that the intrusion signal is a past copy of the original output. A loop transmission of such signal to the real system constitutes a stealthy and successful attack, that can be detected by a feedback controller \cref{eq: controlform} that includes the term \cref{eq: controlformdelay}. 
\subsection{The $\chi^2$ Fault Detector}
The availabel data to examine are the observed output \cref{output_eqn2} and the Kalman estimate \cref{eq: kalman}. The $\chi^2$ detector is a standard tool in system diagnostics that leverages these data to detect data incoherencies that can be measured from the residuals $y-C\hat{x}$, \cite{mehra1971innovations}. The next result characterizes the limit distribution of the residuals.
\begin{theorem}\label{thm: res}\cite{mehra1971innovations}
For system \cref{state_eqn2,output_eqn2} with filter \cref{eq: kalman,eq: kalmanoutput} and LQG controller \cref{eq: controlformlqg}, the residues $y_t-C\hat{x}_{t|t-1}$ are IID Gaussian distributed for every $t$, with $\lim_{t} y_t-C\hat{x}_{t|t-1} \sim \mathcal N(0,\Sigma_{R}:=C P C^T +\Sigma_V)$. 
\end{theorem} Given a time detection window $T$, the $\chi^2$ detector within $T$ is defined as 
\begin{equation}\label{eq: chi}
    g_{\kappa}(T) = \sum_{t=\kappa}^{\kappa+T} (y_t-C\hat{x}_{t})^T \Sigma_{R}^{-1} (y_t-C\hat{x}_{t})
\end{equation}
In view of Theorem \ref{thm: res}, $g_{\kappa}(T)$ follows a $\chi^2$ distribution with $Tn_y$ degrees of freedom \cite{scharf1991statistical}. 
Manipulation of \eqref{eq: chi} with \eqref{eq: clsystem} leads to a  crucial observation: in the absence of replay attacks the distribution of residual is agnostic to watermarking statistics (i.e. randomness induced by $u_t^{WM}$). Thus Theorem \ref{thm: res} is valid when the system is not under attack. In the remainder of the section we will explore the effect of \cref{eq: controlformlqg} on $\mathbb E[g_{\kappa}(T)].$

\subsection{Detection Analysis}
During attacks, the compromised readings $y'(t)$, modify the $\chi^2$ detector to: 
\begin{equation}\label{eq: chiAttack}
    g'_\kappa(T)  = \sum_{t=\kappa}^{\kappa+T} (y'_t-C\hat{x}_{t})^T \Sigma_{R}^{-1} (y'_t-C\hat{x}_{t}).
\end{equation}
Note that for $\kappa>>1$ the expected value of \cref{eq: chi} or \cref{eq: chiAttack}  for is calculated by considering the expected value of the sum elements when $t\rightarrow +\infty$. In particular for \cref{eq: chiAttack}  we observe that 
\begin{align*}
    &\nonumber \lim _{t \rightarrow + \infty} \mathbb E\left[\left(y_{t}^{\prime}-C \hat{x}_{t}\right)^{T} \Sigma_{R}^{-1}\left(y_{t}^{\prime}-C \hat{x}_{t}\right)\right] 
    \\& =\operatorname{trace}\left[\big(\lim _{t \rightarrow \infty} \operatorname{Var}\left(x_{t}^{\prime}-\hat{x}_{t}\right) + \Sigma_V  \big)  C^T\Sigma_{R}^{-1} C\, \right]
\end{align*}
 Following \cite{mo2009secure}, we write the residual element of time $t$ from \eqref{eq: chiAttack} as $ x'_t-\hat{x}_{t} =  \big(x'_t- \hat{x}'_{t}\big) +  \big(\hat{x}'_{t} - \hat{x}_{t} \big)$.
for which the variance yields
\begin{equation*}
\begin{split}
    \operatorname{Var} \big( x'_t-\hat{x}_{t}\big) =  & \operatorname{Var}  \big (x'_t- \hat{x}'_{t} \big) + \operatorname{Var}  \big(\hat{x}'_{t} - \hat{x}_{t} \big) + \\ &\operatorname{Cov}\big(x'_t- \hat{x}'_{t} , \hat{x}'_{t} - \hat{x}_{t}  \big) + \\ & \operatorname{Cov}\big( \hat{x}'_{t} - \hat{x}_{t} , x'_t- \hat{x}'_{t}  \big)
    \end{split}
\end{equation*} The first term comes exclusively from the virtual system and from the discussion above it can be deduced that  
$ \operatorname{trace}\left[\big( \operatorname{Var}  (x'_t- \hat{x}'_{t} ) + \Sigma_{V}\big) C^{T}\Sigma_{R}^{-1}C \right] \rightarrow Tn_y $. The rest of the terms can be represented with the use of \eqref{eq: ssautocovariance} and projection operators 
$\boldsymbol{\mathfrak P}\,\mathbb{X}_t : = \hat{x}_{t}' - \hat{x}_t$, $ \boldsymbol{\mathfrak Q}\,\mathbb{X}_t : =  x_{t}' - \hat{x}_t'$. Straightforward algebra yields
\begin{equation}\label{eq: attacksteadysteadyresidual}
\begin{split}
    &\lim_{\kappa\rightarrow +\infty} \mathbb E\big[g'_{\kappa}(T)\big]=Tn_y+\\
    &~~~+T\operatorname{trace}\left[  \boldsymbol{\big(\mathfrak P} \mathcal{C}  \boldsymbol{\mathfrak P}^T+2\boldsymbol{\mathfrak P} \mathcal{C}\boldsymbol{\mathfrak Q}^T \big)\,C^{T}\Sigma_{R}^{-1}C \right].
\end{split}
\end{equation}

\subsection{Practical Considerations} Formula \eqref{eq: attacksteadysteadyresidual} demonstrates that, when system transitions from clean to attacked, the $\chi^2$-detector gets triggered so that $\lim_{\kappa\rightarrow +\infty} \mathbb E\big[g'_{\kappa}(T)\big]>\lim_{\kappa \rightarrow +\infty} \mathbb E\big[g_{\kappa} (T) \big] = T n_y$, and \eqref{eq: attacksteadysteadyresidual} quantifies the difference. A strong enough deviation of $\mathbb E[g_{\kappa}(T)]$ should be a reliable attack detector. The practical way that $g_{\kappa}(T)$ is utilized for fault detection is by comparing the sample mean against a threshold $\psi : = \mathcal O ( T n_y ) $. The intuition is that in the absence of an attack the residuals' statistics are as in Theorem \ref{thm: res} and $\eta$ can be the cutoff of false positive attack. In the attack event, the compromised signal $y_t$, denoted as $y'_t$, gets compared against $C \hat{x}$ must yield different statistics from Theorem \ref{thm: res}. 

\section{ Performance Implications }\label{sec: time-delayperformance}
The feedback controller is designed to minimize Eq. \eqref{eq: lqg}. An additive input control signal will yields in higher cost, i.e. lower system performance. The effect of \eqref{eq: controlformdelay} in the value of \eqref{eq: lqg} occurs as an extra term to the optimal control cost value. Indeed a standard Dynamic Programming argument shows that \eqref{eq: lqg} under control \eqref{eq: controlform}, denoted as $J_{\text{WM}}$, can be cast as 
\begin{equation}\label{eq: newcost}
   J_{\text{WM}} = J_{*} + \operatorname{trace}\big[ K_{\tau}^T (B^T\,\Delta_0\,B+R) K_{\tau} \big].
\end{equation} where $J_*$ is the optimal cost value of \eqref{eq: lqg} with \eqref{eq: controlformlqg} only, and $\Delta_o=\Delta_o^T$ the solution of the Ricatti equation
\begin{equation*}
    \Delta_o = Q + A^T \Delta_o\, A - A^T \Delta_o \,B (R+B^T\Delta_o\,B)^{-1}B^T\Delta_o\, A.
\end{equation*}
The steps to derive \eqref{eq: newcost} are similar to cost expressions derived in \cite{mo2009secure}.
\section{Simulation Example} \label{sec:sim} 
\begin{figure}[t]
    \centering
    \includegraphics[scale=0.6]{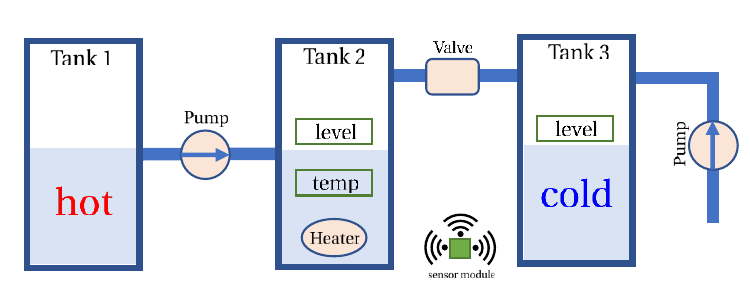}
    \caption{\small{The chemical process with four actuators (hot, cold pumps, valve, and heater) that control the level and temperature of tank 2 and level of tank 3.}}
    \label{fig: tank}
\end{figure}

We consider a chemical process shown in Figure~\ref{fig: tank} linearized along the lines of \cite{milovsevic2019estimating}. The control inputs are two flow pumps, as illustrated in Figure \ref{fig: tank}. The three states $x=(x_1,x_2,x_3)^T$ are the level of water in tanks 2 and 3 and the temperature of water in tank 2. The process and measurement noises are zero-mean Gaussian with covariances $W$$=$$0.5\,I_{n_x}$ and $V$$=$$0.1\,I_{n_y}$, respectively. The dynamics of this three-tank system are
$$A\hspace{-0.05in}  = \hspace{-0.05in} \begin{bmatrix}
0.96 & 0 & 0 \\
0.04 & 0.97 & 0 \\
-0.04 & 0 & 0.9
\end{bmatrix},
B \hspace{-0.05in} = \hspace{-0.05in} \begin{bmatrix}
8.8 & -2.3 & 0 & 0\\
0.2 & 2.2 & 4.9 & 0 \\
-0.21 & -2.2 & 1.9 & 21
\end{bmatrix}$$
and also $C=I_{n_x}$.  We choose the time window as $T = 85$, threshold $\psi = 110$, the time-delays take values in $[50,200]$ and time-delay feedback matrix $K_{\tau} =  0.0713 I_{n_x}$, designed to meet Theorem \ref{thm: stability}. The performance metric to measure the system response is considered as the LQG cost with the weighting matrices: $Q_s = \text{diag}[0.3,0.3,2.4]$ for state cost, and $R = \text{diag}[1,1,1,1]$ for input cost.

\begin{figure}[ht!]
    \centering
    \includegraphics[scale=0.45]{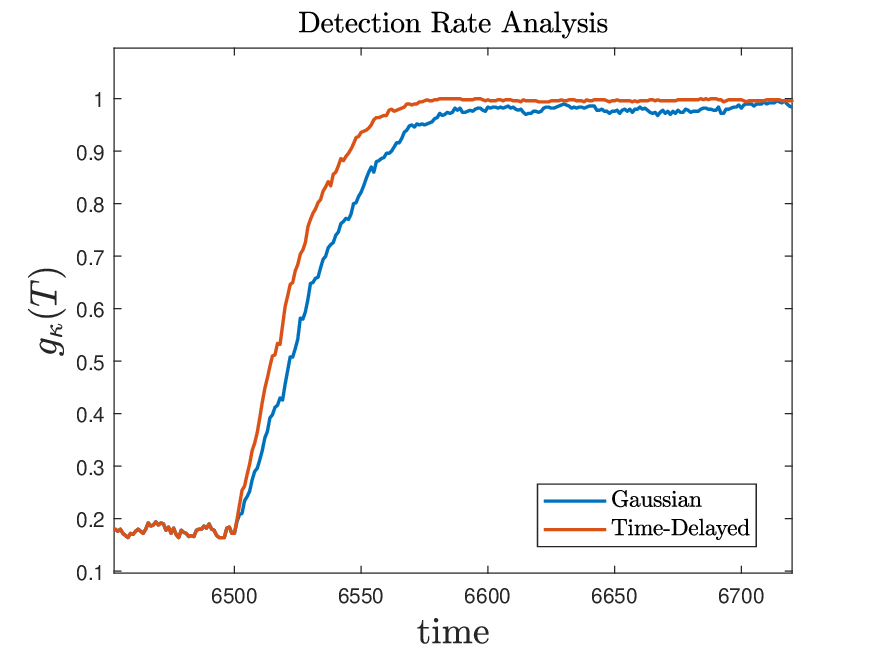}
    \caption{\small{Comparison of the detection rate of a replay attack between additive Gaussian and delay-induced feedback watermarking. Attack event starts at time $t'=6500$.}}
    \label{fig:comparison}
\end{figure}
We compare detection rates between the time-delayed watermarking, and an additive Gaussian watermarking signal. The parameters for the Gaussian watermarking signal are chosen to be zero mean with a stationary covariance of $\Sigma_{GW} = 0.015 I_{n_x}$. We ensure a fair comparison by making the detection rate comparable to that of the time-delayed watermarking. Figure \ref{fig:comparison} shows the results for the detection rate for the attacked system for both Gaussian watermarking and the time-delay watermarking. In the simulations, we record the output of the system from time $t_{start} = 6000$ to $t_{end} = 6300$, and the replay the output values starting from time $t'= 6500$. The steps change of the calculated $\chi^2$ measure $g_{\kappa}(T)$ past $t'$ highlights that some sort of attack is taking place. The time-delayed watermarking signal results in a faster and overall better detection rate. Finally, the system performance without watermarking (optimal) is 0.7907, with Additive Gaussian is 1.0415 and with Time-Delayed watermarking is 0.8712. 
%

\section{Discussion}\label{sec: discussion}
We presented a new type of watermarked security in linear systems. We leveraged a type of random time-delayed feedback for detection of replay attacks. The theoretical framework poses several challenges and key points to pay attention to for the effective synthesis of watermarking signals.  The working hypothesis that motivated this effort relies on the advantages of state-feedback control for fault and attack detection in the closed-loop system performance, over additive noise. 

\appendix \label{append: matrices}
%
$$
    \mathcal Q = \begin{bmatrix}
        \Sigma_W &  O &  O &  O \\
        O^T & \Sigma_V & O_{n_v} &  O_{n_v}  \\
        O^T & O_{n_v}  & \Sigma_V  & \big(\sum_{\tau=1}^{\overline{\tau}}p^2_\tau\big) \Sigma_V \\
        O^T & O_{n_v} & \big(\sum_{\tau=1}^{\overline{\tau}}p^2_\tau\big) \Sigma_V  & \Sigma_V 
    \end{bmatrix}
$$ 
$$
    \mathcal{R}_{l,\tau,\tau'} = \begin{bmatrix}
     O_{n_w} &  O  & O &  O \\
     O^T & O_{n_v} & O_{n_v} & O_{n_v}  \\
     O^T & O_{n_v}  & p_{l+\tau} \Sigma_V & p_{l+\tau} \Sigma_V \\
     O^T & O_{n_v}  & p_{l+\tau'} \Sigma_V & p_{l+\tau'} \Sigma_V
\end{bmatrix}
$$ 
$$\mathcal S =\begin{bmatrix}
O_{n_w} & O & O & O\\
O^T & O_{n_v} & O_{n_v}  & O_{n_v} \\
O^T & \Sigma_V & O_{n_v} & O_{n_v} \\
O^T & \Sigma_V  & O_{n_v} & O_{n_v}
\end{bmatrix} $$
with $O=O_{n_w\times n_v}$.
\bibliographystyle{IEEEtran}

\bibliography{refs}

\begin{thebibliography}{10}
\providecommand{\url}[1]{#1}
\csname url@samestyle\endcsname
\providecommand{\newblock}{\relax}
\providecommand{\bibinfo}[2]{#2}
\providecommand{\BIBentrySTDinterwordspacing}{\spaceskip=0pt\relax}
\providecommand{\BIBentryALTinterwordstretchfactor}{4}
\providecommand{\BIBentryALTinterwordspacing}{\spaceskip=\fontdimen2\font plus
\BIBentryALTinterwordstretchfactor\fontdimen3\font minus \fontdimen4\font\relax}
\providecommand{\BIBforeignlanguage}[2]{{%
\expandafter\ifx\csname l@#1\endcsname\relax
\typeout{** WARNING: IEEEtran.bst: No hyphenation pattern has been}%
\typeout{** loaded for the language `#1'. Using the pattern for}%
\typeout{** the default language instead.}%
\else
\language=\csname l@#1\endcsname
\fi
#2}}
\providecommand{\BIBdecl}{\relax}
\BIBdecl

\bibitem{VENKATASUBRAMANIAN2003293}
V.~Venkatasubramanian, R.~Rengaswamy, K.~Yin, and S.~N. Kavuri, ``A review of process fault detection and diagnosis: Part i: Quantitative model-based methods,'' \emph{Computers \& Chemical Engineering}, vol.~27, no.~3, pp. 293--311, 2003.

\bibitem{SurveyReplay}
H.~Liu, Y.~Mo, and K.~H. Johansson, ``Active detection against replay attack: A survey on watermark design for cyber-physical systems,'' in \emph{Safety, Security and Privacy for Cyber-Physical Systems}.\hskip 1em plus 0.5em minus 0.4em\relax Springer, 2021, pp. 145--171.

\bibitem{sharma2010sensor}
A.~B. Sharma, L.~Golubchik, and R.~Govindan, ``Sensor faults: Detection methods and prevalence in real-world datasets,'' \emph{ACM Transactions on Sensor Networks (TOSN)}, vol.~6, no.~3, 2010.

\bibitem{cardenas2008secure}
A.~A. Cardenas, S.~Amin, and S.~Sastry, ``Secure control: Towards survivable cyber-physical systems,'' in \emph{2008 The 28th International Conference on Distributed Computing Systems Workshops}.\hskip 1em plus 0.5em minus 0.4em\relax IEEE, 2008, pp. 495--500.

\bibitem{rubio2017use}
J.~Rubio-Hernan, L.~De~Cicco, and J.~Garcia-Alfaro, ``On the use of watermark-based schemes to detect cyber-physical attacks,'' \emph{EURASIP Journal on Information Security}, vol. 2017, no.~1, pp. 1--25, 2017.

\bibitem{mehra1971innovations}
R.~K. Mehra and J.~Peschon, ``An innovations approach to fault detection and diagnosis in dynamic systems,'' \emph{Automatica}, vol.~7, no.~5, pp. 637--640, 1971.

\bibitem{mo2009secure}
Y.~Mo and B.~Sinopoli, ``Secure control against replay attacks,'' in \emph{2009 47th annual Allerton conference on communication, control, and computing (Allerton)}.\hskip 1em plus 0.5em minus 0.4em\relax IEEE, 2009, pp. 911--918.

\bibitem{astrom}
K.~{\AA}str{\"o}m, \emph{Introduction to Stochastic Control Theory}, ser. Dover Books on Electrical Engineering.\hskip 1em plus 0.5em minus 0.4em\relax Dover Publications, 2006.

\bibitem{scharf1991statistical}
L.~Scharf and C.~Demeure, \emph{Statistical Signal Processing: Detection, Estimation, and Time Series Analysis}, ser. Addison-Wesley series in electrical and computer engineering.\hskip 1em plus 0.5em minus 0.4em\relax Addison-Wesley Publishing Company, 1991.

\bibitem{milovsevic2019estimating}
J.~Milo{\v{s}}evi{\'c}, H.~Sandberg, and K.~H. Johansson, ``Estimating the impact of cyber-attack strategies for stochastic networked control systems,'' \emph{IEEE Transactions on Control of Network Systems}, vol.~7, no.~2, pp. 747--757, 2019.

\end{thebibliography}

\end{document}